\pdfoutput=1

\documentclass[11pt]{article}

\usepackage[final]{acl}

\usepackage{times}
\usepackage{latexsym}

\usepackage[T1]{fontenc}

\usepackage[utf8]{inputenc}

\usepackage{microtype}

\usepackage{inconsolata}

\usepackage{graphicx}
\usepackage{booktabs}
\usepackage{amsmath}
\usepackage{amssymb}
\usepackage{multirow}
\usepackage{enumitem}
\usepackage{float}  
\title{HetGCoT: Heterogeneous Graph-Enhanced Chain-of-Thought LLM Reasoning for Academic Question Answering}

\author{
Runsong Jia\textsuperscript{1}, 
Mengjia Wu\textsuperscript{1}, 
Ying Ding\textsuperscript{2}, 
Jie Lu\textsuperscript{1}, 
Yi Zhang\textsuperscript{1} \\
\\
\textsuperscript{1}University of Technology Sydney, Sydney, Australia \\
\textsuperscript{2}University of Texas at Austin, Austin, United States \\
\\
\texttt{\{runsong.jia, mengjia.wu, jie.lu, yi.zhang\}@uts.edu.au} \\
\texttt{ying.ding@ischool.utexas.edu}
}

\begin{document}
\maketitle
\begin{abstract}
Academic question answering (QA) in heterogeneous scholarly networks presents unique challenges requiring both structural understanding and interpretable reasoning. While graph neural networks (GNNs) capture structured graph information and large language models (LLMs) demonstrate strong capabilities in semantic comprehension, current approaches lack integration at the reasoning level. We propose HetGCoT, a framework enabling LLMs to effectively leverage and learn information from graphs to reason interpretable academic QA results. Our framework introduces three technical contributions: (1) a framework that transforms heterogeneous graph structural information into LLM-processable reasoning chains, (2) an adaptive metapath selection mechanism identifying relevant subgraphs for specific queries, and (3) a multi-step reasoning strategy systematically incorporating graph contexts into the reasoning process. Experiments on OpenAlex and DBLP datasets show our approach outperforms all sota baselines. The framework demonstrates adaptability across different LLM architectures and applicability to various scholarly question answering tasks.

\end{abstract}

\section{Introduction}
Academic question answering in heterogeneous scholarly networks presents essential challenges in integrating structural knowledge with semantic understanding. QA tasks regarding publishing venue selection, paper authorship, and scientific collaboration all require systems to reason over complex networks of papers, authors, venues, and organizations while providing interpretable explanations \cite{shi2018heterogeneous, ji2021survey}.

The academic knowledge space is inherently heterogeneous, comprising diverse entities (e.g., papers, authors, venues and organizations) connected through various relationship types. Effective academic question answering systems must address three fundamental challenges: (1) modeling heterogeneous structures to capture complex relationships across different entity types and query contexts, (2) adaptively selecting relevant knowledge subgraphs based on query semantics rather than uniformly processing entire network structures, and (3) transforming structural knowledge into coherent natural language explanations that can justify answers across different academic QA scenarios. While these challenges manifest differently across different tasks, they share the common requirement of integrating graph-structured knowledge with semantic reasoning.

Current approaches to academic question answering have attempted to address these challenges through various strategies. However, existing methods face significant limitations in addressing these challenges holistically. Heterogeneous graph neural networks (HGNNs) can effectively model complex academic networks \cite{hu2020heterogeneous}, but struggle with: (1) adapting their representations to different query types and relationship patterns, (2) generating task-specific subgraph selections, and (3) producing natural language explanations for diverse academic QA scenarios. LLMs demonstrate strong semantic understanding \cite{chowdhery2022palm} but cannot directly process the rich structural information embedded in academic networks. Existing integration attempts typically focus on single tasks or treat graph information as auxiliary features through simple concatenation, failing to systematically incorporate structural patterns into the reasoning process across diverse academic QA scenarios \cite{zhao2023unifying}.

To address these limitations, we propose HetGCoT (Heterogeneous Graph-Enhanced Chain-of-Thought), a framework that integrates heterogeneous graph neural networks with large language models for academic question answering. HetGCoT transforms graph structural patterns into confidence-weighted natural language reasoning chains through metapath naturalization, enabling LLMs to process complex academic relationships. The framework employs adaptive metapath selection using Heterogeneous Graph Transformer (HGT) \cite{hu2020heterogeneous} embeddings and FastGTN-learned \cite{yao2021fastgtn} importance weights to dynamically identify task-relevant subgraphs. Through a multistep chain-of-thought reasoning process, HetGCoT anchors on three task-driven analytical foci: analyzing venue patterns for journal recommendation, temporal relationships for authorship queries, and collaboration networks for collaboration discovery. This integrated approach enables deep reasoning-level fusion of graph structures with language understanding across diverse academic QA scenarios.

Through extensive experiments on OpenAlex and DBLP datasets \cite{priem2022arxiv}, we demonstrate HetGCoT's effectiveness across multiple academic QA tasks. For journal recommendation, our framework achieves 92.21\% and 83.70\% H@1 accuracy respectively. Moreover, we validate its generalizability on historical publication QA (author-paper reasoning) and author collaboration QA (author-paper-author reasoning), showing consistent improvements on general academic QA tasks. 

The key contributions of this work include:

\begin{itemize}[leftmargin=1em, itemsep=\parskip]
\item A unified framework for academic question answering that transforms heterogeneous graph structures into LLM-processable reasoning chains, enabling effective integration of structural and semantic understanding for academic question answering
\item An adaptive metapath selection mechanism that dynamically identifies relevant subgraphs based on query characteristics, supporting various academic QA scenarios
\item A flexible multi-step reasoning strategy that adapts to different academic QA tasks while maintaining systematic integration of graph-derived contexts
\end{itemize}

\section{Related Works}

\subsection{LLMs and Reasoning}
LLMs have revolutionized natural language processing through their sophisticated understanding and generation capabilities. Building upon the Transformer architecture \citep{vaswani2017attention}, prominent models including GPT \citep{brown2020language}, LLaMA \citep{touvron2023llama}, Qwen \citep{qwen2023}, and PaLM \citep{chowdhery2022palm} have achieved remarkable performance across diverse language tasks.

A pivotal advancement is chain-of-thought (CoT) reasoning \citep{wei2022chain}, which enhances LLMs' ability to tackle complex problems through explicit intermediate reasoning steps. This approach has proven particularly effective for tasks requiring multi-step inference and logical decomposition. Extensions such as self-consistency \citep{wang2023self} and tree-of-thought \citep{yao2023tree} further refine this capability, establishing structured reasoning frameworks for specialized domains.

\subsection{Integration of GNNs and LLMs}
The integration of graph neural networks with language models has emerged as a promising direction for leveraging both structural and semantic information. Recent work explores various integration strategies to combine the complementary strengths of both modalities.

\textbf{Graph Prompting and Reasoning Methods:} Several frameworks attempt to enhance LLMs with graph-based reasoning. GraphPrompter \citep{liu2024soft} explore soft prompting techniques for graph learning tasks with LLMs. Graph Chain-of-Thought \citep{jin2024graphcot} augments LLMs by explicitly reasoning on graph structures, while Graph of Thoughts \citep{besta2024graph} models the reasoning process itself as a graph structure. Think-on-Graph \citep{sun2023think} proposes deep reasoning executed directly on knowledge graphs.

\textbf{Retrieval-Augmented Approaches:} PathRAG \citep{chen2025pathrag} enhances LLMs through graph-based retrieval using relational paths, while GNN-RAG \citep{mavromatis2024gnnrag} combines graph neural retrieval with language model reasoning. Generate-on-Graph \citep{chen2024generate} treats LLMs as both agents and knowledge graphs for incomplete QA tasks.

\textbf{Heterogeneous Graph and Metapath Methods:} For academic networks specifically, heterogeneous graph neural networks like HGT \citep{hu2020heterogeneous} and HAN \citep{wang2019heterogeneous} model complex relationships between different entity types. Metapath-based techniques provide interpretable relationship modeling through typed connection sequences. Recent work such as Metapath of Thoughts \citep{solanki2024metapath} verbalizes metapaths as contextual augmentation for LLMs. While these methods show promise, they typically focus on single tasks or treat graph information as auxiliary features rather than achieving deep reasoning-level integration.

Despite these advances, existing approaches face limitations in: (1) adaptively selecting task-relevant subgraphs, (2) transforming heterogeneous structural patterns into natural language reasoning chains, and (3) systematically integrating graph-derived contexts throughout the reasoning process. Our HetGCoT framework addresses these gaps by introducing adaptive metapath selection with learned importance weights, metapath naturalization for LLM processing, and a structured multi-step reasoning strategy that deeply integrates graph knowledge at each reasoning stage.

\section{Methodology}

\begin{figure*}[t]
  \centering
  \includegraphics[width=\textwidth]{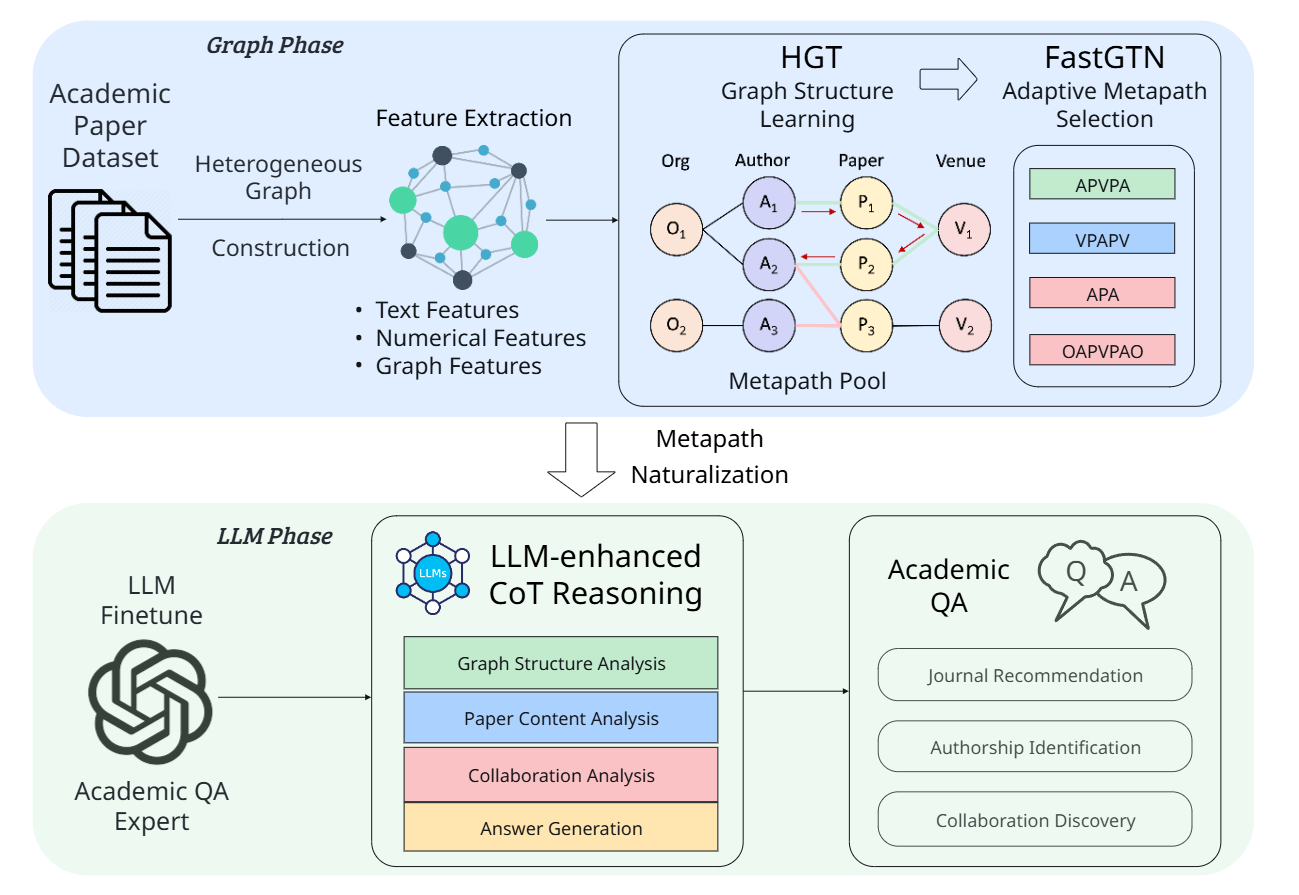}
  \caption{HetGCoT Overall architecture.}
  \label{fig:architecture}
\end{figure*}

In this section, we present our proposed HetGCoT framework. Figure~\ref{fig:architecture} illustrates the system architecture designed to address academic question answering through the integration of heterogeneous graph structural information with LLM reasoning capabilities.

We consider a heterogeneous academic graph \( G = (V, E, \phi, \psi) \), where \( V = V_p \cup V_a \cup V_v \) represents the set of nodes comprising papers (\( V_p \)), authors (\( V_a \)), and venues (\( V_v \)). \( E \) denotes the set of edges \( E = E_{PV} \cup E_{PA} \), capturing paper-venue and paper-author relationships, with \( \phi: V \to A \) mapping nodes to their types and \( \psi: E \to R \) mapping edges to their relationship types. Given a query $q$ (which could be a paper, author, or research topic), our task is to provide accurate answers with interpretable explanations.

\subsection{Heterogeneous Academic Graph Construction}
We construct a heterogeneous academic graph with three node types (papers, authors, venues) and two edge types (paper-venue, paper-author). Node features $\mathbf{x}_v$ are initialized through:
\begin{equation}
    x_v = \text{LayerNorm}(\text{concat}(x_{\text{text}}, x_{\text{num}}))
\end{equation}
where $x_{\text{text}}$ are Sentence-BERT \citep{reimers2019sentence} encoded titles, abstracts, and keywords, and $x_{\text{num}}$ include citation counts, impact factors, and other numerical attributes. This framework can extend to additional node and edge types for different academic QA tasks. This comprehensive feature engineering ensures our model can leverage both content semantics and academic impact signals.

We then employ HGT to encode graph structure. HGT is particularly suitable for academic networks where nodes and relationships naturally possess varying semantic importance. Unlike traditional GNNs that treat all nodes homogeneously, Unlike traditional GNNs that treat all nodes homogeneously, HGT incorporates type-aware attention mechanisms that effectively capture this heterogeneity, allowing the model to differentiate between various node and edge types through specialized attention calculations:
\begin{equation}
    \mathbf{h}_i^{(l)} = \sum_{j \in \mathcal{N}(i)} \sum_{r \in \mathcal{R}} \alpha_{i,j,r}^{(l)} \cdot \mathbf{W}_r^{(l)}\mathbf{h}_j^{(l-1)}
\end{equation}
where $\mathbf{h}_i^{(l)}$ represents the $l$-th layer embedding of node $i$, $\mathcal{N}(i)$ denotes its neighbors, $\mathcal{R}$ is the set of relation types, $\alpha_{i,j,r}^{(l)}$ are type-aware attention weights, and $\mathbf{W}_r^{(l)}$ are relation-specific transformation matrices. This encoding captures the semantic importance variations essential for subsequent metapath selection.

The model is trained with a link prediction objective tailored to the target task. The output embeddings encode both local neighborhood information and global patterns, forming the basis for metapath selection.

\subsection{Adaptive Metapath Selection}
We leverage metapaths to capture structured evidence in heterogeneous academic networks. Each metapath $\pi$ represents a sequence of relations connecting different node types.

We define four metapath templates: (1) \textbf{APVPA} capturing venue-based author connections, (2) \textbf{VPAPV} identifying venue relationships through shared authors, (3) \textbf{APA} representing direct collaborations, and (4) \textbf{OAPVPAO} capturing institutional connections. These templates, inspired by heterogeneous network embedding approaches \cite{dong2017metapath2vec}, comprehensively capture the semantic structures in academic heterogeneous graphs, providing rich relational contexts for academic QA tasks.

For each query node, we first generate a candidate pool of metapath instances by identifying semantically similar entities using cosine similarity of HGT embeddings as semantic starting points. This approach leverages the encoded structural representations to identify relevant subgraphs for exploration.

Unlike traditional approaches using manually defined importance, we employ FastGTN to learn relationship importance weights automatically:
\begin{equation}
\mathbf{H}^{(l,c)} = \sigma\left(\sum_{r=1}^{|\mathcal{R}|} \alpha_{l,r}^{(c)} \mathbf{A}^{(r)}\mathbf{H}^{(l-1,c)}\mathbf{W}^{(l,c)}\right)
\end{equation}

where $\mathbf{H}^{(l,c)}$ denotes the representation at the $l$-th layer in channel $c$, $\alpha_{l,r}^{(c)}$ are the learned relation importance weights crucial for metapath scoring, $\mathbf{A}^{(r)}$ represents the adjacency matrix for relation type $r$, and $\sigma$ is the activation function.

We train FastGTN with a self-encoding objective, minimizing reconstruction error for paper nodes. This approach offers two key advantages: it requires no additional labeling, enabling fully unsupervised learning of relation importance; and it forces the model to identify which relation combinations best preserve node semantics. Notably, we repurpose FastGTN as an explanation generator rather than a classifier, extracting relation importance weights that quantify the semantic significance of different metapaths. Importantly, we use frozen HGT embeddings as input features to FastGTN, ensuring complementarity between the two models: HGT provides node-level semantic embeddings, while FastGTN discovers global relation patterns.

After training, we extract relation importance weights from the model by averaging weights across all layers and channels. We then score each metapath instance by summing the learned importance weights of its constituent edges:
\begin{equation}
\text{score}_{\text{norm}}(\pi) = \frac{\sum_{(u,v) \in \pi} w_{\psi(u,v)}}{|\pi|^{\gamma}}
\end{equation}

where $\pi$ denotes a metapath instance, $w_{\psi(u,v)}$ represents the FastGTN-learned weight for edge $(u,v)$ of type $\psi(u,v)$, and $\gamma \in [0, 1]$ controls length normalization, with larger values increasingly penalizing longer paths.

We employ a stratified selection strategy, taking the top-\textit{k} paths from each template to ensure structural diversity rather than global ranking. This approach guarantees that all semantic templates are represented, the highest quality instances within each category are selected, and no single template dominates due to higher absolute scores. We empirically set \textit{k}=5 to optimize the trade-off between contextual richness and prompt manageability.

\subsection{Metapath Naturalization and Chain-of-Thought Enhanced Academic Reasoning}

\paragraph{Metapath Naturalization}
To bridge the gap between graph structure and language models, we transform the selected metapaths into natural language descriptions. This transformation follows a template-based approach, where each metapath type is associated with a specific language template that captures its semantic meaning. Each natural language description is prefixed with a confidence score derived from the FastGTN path scoring mechanism, allowing the LLM to weigh structural evidence according to its reliability. This naturalization process converts graph structural patterns into coherent textual contexts that LLMs can effectively process and reason about.

\paragraph{Multi-step Reasoning Framework}
We design a structured four-step reasoning framework that systematically integrates graph-derived information with content analysis. This CoT approach follows the cognitive process adaptable to different academic QA tasks:

1. \textbf{\textit{Graph Structure Analysis}}: The model processes naturalized metapath (VPAPV, APVPA) information to understand structural patterns in the academic network, focusing on relationship evidence pointing to potential answers.

2. \textbf{\textit{Content Analysis}}: Examines the target paper's specific information (title, abstract, keywords, citation metrics) to identify thematic alignment with candidate answers.

3. \textbf{\textit{Collaboration Analysis}}: Analyzes author collaboration patterns using author-centric metapaths (APA, OAPVPAO) to identify research communities and publication preferences.

4. \textbf{\textit{Answer Generation}}: Synthesizes insights from the previous steps to generate answers with comprehensive explanations.

Each reasoning step receives specifically tailored input information and questions that guide the reasoning process. This structured decomposition improves reasoning transparency while maintaining adaptability across different academic QA scenarios.

\subsection{LLM Enhancement}
We enhance the LLM's reasoning capabilities through prompt engineering and task-specific fine-tuning. The prompt template includes a system message defining the model's role as an academic QA expert and establishing task-relevant constraints. The user message structures input according to our four-step reasoning process.

We fine-tune the model (GPT-4o mini) on datasets containing structured reasoning examples across academic QA tasks. During fine-tuning, we optimize the probability of generating correct answers conditioned on both query semantics and graph-derived contexts:

\begin{equation}
\mathcal{L} = \arg\max_\theta \log P(a|q, \mathcal{M}_s; \theta)
\end{equation}

\begin{equation}
\begin{split}
\mathcal{L} &= \arg\max_\theta \log \sum_{m \in \mathcal{M}_s} \text{score}_{\text{norm}}(m) \\
&\quad \quad \cdot P(a|q, \text{naturalize}(m); \theta)
\end{split}
\end{equation}

where $a$ denotes the target answer, $q$ represents the input query, $\mathcal{M}_s$ is the set of selected metapaths, $\text{score}_{\text{norm}}(m)$ are the FastGTN-learned confidence weights from Equation (4), and $\text{naturalize}(m)$ transforms metapath $m$ into natural language context for LLM processing.

This process teaches the model to: (1) interpret naturalized metapaths as structural evidence, (2) extract relevant information from multiple sources, and (3) generate evidence-supported answers connecting structural patterns with semantic understanding.

During inference, we construct task-specific prompts incorporating adaptive metapath information and query details. This integration creates transparency in the answering process, providing users with clear explanations grounded in both network structure and content semantics.

\section{Experiments}
\subsection{Experimental Setup}
\paragraph{Datasets}
We evaluate the proposed \textbf{HetGCoT} framework on two academic datasets,
\textbf{OpenAlex} and \textbf{DBLP}.
To ensure paper quality and the interpretability of results, we extracted a random subgraph from OpenAlex limited to journals ranked "A" or higher according to the CORE list (a widely used venue ranking system for computing research primarily in Australia and New Zealand), while for DBLP, we randomly sampled a subgraph from the entire dataset. We train the models separately on these two datasets. Table~\ref{tab:datasets} summarises the statistics of the heterogeneous graphs extracted for our experiments.
For each node type, we retain the following attributes:
\emph{venue} (type, name);
\emph{paper} (type, keywords, abstract, citations, FWCI (field-weighted citation impact), title, year);
\emph{author} (type, organization, name).
All textual fields are encoded using Sentence‑BERT, yielding initial node representations that
capture semantic content in titles, abstracts, and keywords.

\begin{table}[t]
\caption{Statistics of experimental datasets.}
\label{tab:datasets}
\centering
\begin{tabular}{lcc}
\toprule
\textbf{Feature} & \textbf{OpenAlex} & \textbf{DBLP} \\
\midrule
Total Nodes & 76,569 & 62,443 \\
Total Edges & 105,290 & 79,697 \\
\midrule
\textbf{Node Distribution} & & \\
venue & 111 & 51 \\
paper & 22,028 & 17,850 \\
author & 54,430 & 44,542 \\
\midrule
\textbf{Edge Distribution} & & \\
paper-venue & 22,028 & 17,850 \\
paper-author & 83,262 & 61,847 \\
\bottomrule
\end{tabular}
\end{table}

\paragraph{Evaluation Metrics}
We report four metrics:
\textbf{Hit} (\%)—measures whether any of the true answers are found in the generated response, which is typically employed when evaluating LLMs;
\textbf{H@1} (\%)—the accuracy of the top/first predicted answer;
\textbf{F1} (\%)—harmonic mean of precision and recall; and
\textbf{NDCG} (\%)—which weights higher‑ranked correct answers more heavily.

\subsection{Baseline Methods}
We compare HetGCoT with three categories of representative baseline methods:
\setlength{\leftmargini}{1em}
\begin{itemize}
  \item \textbf{Pure GNN Methods}:\\
        \textbf{GCN} \citep{kipf2017ICLR}: Basic graph convolutional network that processes homogeneous graph structures\\
        \textbf{GAT} \citep{velivckovic2018graph}: Graph attention network that captures relative importance between nodes through attention mechanisms\\
        \textbf{HGT} \citep{hu2020heterogeneous}: Heterogeneous Graph Transformer, an architecture designed specifically for heterogeneous graphs
  \item \textbf{Pure LLM Methods}:\\
        \textbf{GPT-4o mini}: Base LLM performance in zero-shot settings\\
        \textbf{GPT-4o mini+CoT}: GPT-4o mini model with Chain-of-thought reasoning\\
        \textbf{LLaMA 3 8b+CoT}: LLaMA 3 8B model with Chain-of-thought reasoning
  \item \textbf{Graph+LLM Integration Methods}:\\
        \textbf{PathRAG} \citep{chen2025pathrag}: Retrieval-augmented reasoning based on graph paths\\
        \textbf{GraphPrompter} \citep{liu2024soft}: Graph-structured prompting framework\\
        \textbf{GraphCoT} \citep{jin2024graphcot}: Integrating graph structural information into chain-of-thought reasoning\\
        \textbf{HiGPT} \citep{tang2024higpt}: Heterogeneous graph language model for graph-based reasoning\\        
        \textbf{Graph of Thoughts} \citep{besta2024graph}: Graph-based thought reasoning framework\\
        \textbf{Think-on-Graph} \citep{sun2023think}: Deep reasoning executed on graph structures

\end{itemize}

We further conduct ablation studies with different LLM sizes:
Qwen-2.5 1.5B, Qwen-2.5 7B, LLaMA‑2 7B, LLaMA‑2 13B, and LLaMA‑3 8B, assessing the framework’s adaptability to varying foundation‑model capacities.

\subsection{Results}
\paragraph{\textit{Results on Journal Recommendation}}
While our framework is designed for general academic QA tasks, we primarily demonstrate its effectiveness through journal recommendation due to its representative complexity and practical importance. Table~\ref{tab:main_results} presents a comprehensive performance comparison of HetGCoT against all baseline methods on the OpenAlex and DBLP datasets.

\begin{table*}[t]
  \caption{Performance comparison of HetGCoT against baseline methods on the academic journal recommendation task.}
  \label{tab:main_results}
  \centering
  \setlength{\tabcolsep}{4pt}
  \resizebox{\textwidth}{!}{%
  \begin{tabular}{llcccccccc}
    \toprule
    & & \multicolumn{4}{c}{\textbf{OpenAlex}} & \multicolumn{4}{c}{\textbf{DBLP}} \\
    \cmidrule(lr){3-6} \cmidrule(lr){7-10}
    \textbf{Category} & \textbf{Method} & \textbf{Hit (\%)} & \textbf{H@1 (\%)} & \textbf{F1 (\%)} & \textbf{NDCG (\%)} & \textbf{Hit (\%)} & \textbf{H@1 (\%)} & \textbf{F1 (\%)} & \textbf{NDCG (\%)} \\
    \midrule
    \multirow{3}{*}{Pure GNN}
        & GCN   & 59.49 & 13.92 & 11.82 & 57.08 & 49.84 & 12.48 & 10.85 & 53.29 \\
        & GAT   & 60.14 & 20.68 & 16.02 & 55.54 & 58.28 & 18.39 & 12.56 & 55.54 \\
        & HGT   & 65.83 & 22.36 & 17.20 & 60.59 & 62.58 & 20.72 & 14.59 & 59.70 \\
    \midrule
    \multirow{3}{*}{Pure LLM}
        & GPT-4o mini       & 69.80 & 58.60 & 31.77 & 64.98 & 54.94 & 44.86 & 28.87 & 55.56 \\
        & GPT-4o mini+CoT   & 75.14 & 58.62 & 49.50 & 70.83 & 61.60 & 50.74 & 40.29 & 58.47 \\
        & LLaMA 3 8b+CoT    & 71.23 & 59.21 & 33.64 & 70.72 & 62.67 & 52.79 & 38.47 & 59.68 \\
    \midrule
    \multirow{6}{*}{Graph+LLM}
        & PathRAG          & 76.87 & 66.49 & 35.76 & 75.62 & 67.62 & 63.78 & 49.72 & 61.68 \\
        & GraphPrompter    & 84.83 & 82.68 & 72.37 & 83.79 & 63.64 & 61.63 & 48.65 & 58.28 \\
        & GraphCoT         & 90.47 & 88.25 & 75.39 & 89.59 & 72.79 & 69.62 & 52.67 & 70.72 \\
        & HiGPT            & 90.58 & 88.37 & 76.16 & 87.49 & 81.55 & 79.12 & 60.57 & 78.86 \\
        & Graph of Thoughts& 92.57 & 90.48 & 76.37 & 89.86 & 80.07 & 79.52 & 59.83 & 75.69 \\
        & Think-on-Graph   & 92.85 & 89.27 & 75.36 & 88.36 & 83.75 & 81.89 & 62.76 & 80.68 \\
    \midrule
    \textbf{Ours} & \textbf{HetGCoT} & \textbf{96.48} & \textbf{92.21} & \textbf{79.90} & \textbf{91.29} & \textbf{85.31} & \textbf{83.70} & \textbf{64.55} & \textbf{83.49} \\
    \bottomrule
  \end{tabular}}
\end{table*}
The experimental results reveal several key insights. There is a clear performance improvement trend from pure GNN methods to pure LLM methods to graph+LLM integration methods, indicating the importance of combining structural information with language models. Pure GNN methods (e.g., GCN, GAT, HGT) show limited performance in capturing graph structural information, achieving up to 65.83\% Hit rate and 22.36\% H@1 accuracy. In contrast, pure LLM methods demonstrate stronger capabilities in semantic understanding, reaching 75.14\% Hit rate.

Our HetGCoT framework outperforms all baseline methods across all metrics, achieving 96.48\% Hit rate, 92.21\% H@1, and 79.90\% F1 score on OpenAlex on the journal recommendation task. This improvement can be attributed to our framework's structure-aware mechanism and multi-step reasoning strategy, which effectively integrates heterogeneous graph information with language model reasoning capabilities. Furthermore, HetGCoT maintains strong performance on the DBLP dataset, demonstrating that our method generalizes to different academic data environments.

\begin{table*}[t]
  \caption{Performance on general academic QA tasks.}
  \label{tab:general_results}
  \centering
  \setlength{\tabcolsep}{4pt}
  \resizebox{\textwidth}{!}{%
  \begin{tabular}{llcccccccc}
    \toprule
    & & \multicolumn{4}{c}{\textbf{OpenAlex}}
      & \multicolumn{4}{c}{\textbf{DBLP}}\\
    \cmidrule(lr){3-6}\cmidrule(lr){7-10}
    \textbf{Task} & \textbf{Method}
      & \textbf{Hit (\%)} & \textbf{H@1 (\%)} & \textbf{F1 (\%)} & \textbf{NDCG (\%)}
      & \textbf{Hit (\%)} & \textbf{H@1 (\%)} & \textbf{F1 (\%)} & \textbf{NDCG (\%)}\\
    \midrule
    \multirow{2}{*}{\shortstack[c]{Authorship\\Identification}}
      & Zero-shot & 32.68 & 16.97 & 19.25 & 28.71
                 & 36.62 & 11.07 & 32.21 & 27.16\\
      & HetGCoT   & \textbf{84.42} & \textbf{74.12} & \textbf{82.22} & \textbf{90.00}
                 & \textbf{86.12} & \textbf{75.86} & \textbf{82.71} & \textbf{81.72}\\
    \midrule
    \multirow{2}{*}{\shortstack[c]{Collaboration\\Discovery}}
      & Zero-shot & 22.11 & 6.53  & 7.37  & 15.24
                 & 40.91 & 15.09 & 50.91 & 35.92\\
      & HetGCoT   & \textbf{58.79} & \textbf{29.60} & \textbf{50.91} & \textbf{41.86}
                 & \textbf{67.75} & \textbf{49.11} & \textbf{57.75} & \textbf{51.38}\\
    \bottomrule
  \end{tabular}}
\end{table*}

Moreover, HetGCoT enhances the interpretability of academic QA through its structured reasoning approach. The adaptive metapath selection identifies the most relevant structural evidence for each query, while the four-step reasoning process generates transparent explanations that detail the model's analysis from graph patterns to semantic understanding, providing users with clear rationales for each answer.

\paragraph{\textit{Model Generalization Capability}}
To validate the generalization capability of the HetGCoT framework, we applied it to more general academic question answering tasks beyond journal recommendation, including authorship identification QA (paper-author relationships) and collaboration discovery QA (author-author relationships). 
As shown in Table~\ref{tab:general_results}, HetGCoT consistently improves performance across these tasks. For authorship identification QA, which requires understanding temporal relationships between authors and their publications, HetGCoT demonstrates substantial improvements over the zero-shot baseline across all datasets. Similarly, for the more challenging collaboration discovery QA task, which involves identifying collaboration patterns between authors, our framework delivers notable gains across all metrics. These results indicate that the combination of structure-aware mechanism and multi-step reasoning in HetGCoT effectively adapts to various general academic question answering scenarios.

\paragraph{\textit{Model Adaptability Experiments}}
To verify the plug-and-play nature of the HetGCoT framework and the effect of model scale on performance, we evaluated our method across different-sized LLMs on the Openalex dataset, with results shown in Table~\ref{tab:model_scale}.

\begin{table}[t]
  \caption{Performance comparison of different‑sized LLMs within the HetGCoT framework.}
  \label{tab:model_scale}
  \centering
  \setlength{\tabcolsep}{4pt}
  \resizebox{\columnwidth}{!}{%
  \begin{tabular}{lccc}
    \toprule
    \textbf{Model} & \textbf{Hit (\%)} & \textbf{H@1 (\%)} & \textbf{F1 (\%)} \\
    \midrule
    Qwen-2.5 1.5B zeroshot        & 10.40 & 5.80 & 3.58 \\
    Qwen-2.5 1.5B+HetGCoT         & \textbf{15.33} & \textbf{6.67} & \textbf{4.40} \\
    \midrule
    Qwen-2.5 7B zeroshot          & 47.00 & 46.00 & 14.95 \\
    Qwen-2.5 7B+HetGCoT           & \textbf{78.67} & \textbf{62.67} & \textbf{34.70} \\
    \midrule
    LLaMA-2 7b zeroshot       & 25.34 & 11.29 & 8.98 \\
    LLaMA-2 7b+HetGCoT        & \textbf{38.63} & \textbf{34.28} & \textbf{11.24} \\
    \midrule
    LLaMA-2 13b zeroshot      & 44.68 & 32.78 & 16.32 \\
    LLaMA-2 13b+HetGCoT       & \textbf{59.56} & \textbf{46.48} & \textbf{28.46} \\
    \midrule
    LLaMA-3 8b zeroshot       & 52.47 & 31.46 & 24.59 \\
    LLaMA-3 8b+HetGCoT        & \textbf{75.33} & \textbf{65.33} & \textbf{34.45} \\
    \bottomrule
  \end{tabular}}
\end{table}

Two key trends emerge from the experimental results. First, the HetGCoT framework consistently improves performance across various LLM architectures, from smaller models like Qwen-2.5 1.5B to larger ones such as LLaMA-3 8B, demonstrating its plug-and-play nature. Second, we observe that larger models achieve more substantial gains when enhanced with HetGCoT. For instance, Qwen-2.5 7B improves from a zero-shot Hit rate of 47.00\% to 78.67\%, while smaller models show more modest improvements. This suggests that models with greater capacity can better exploit the heterogeneous graph information provided by our framework.

\subsection{Ablation Study}

\begin{figure}[t]
  \centering
  \includegraphics[width=\columnwidth]{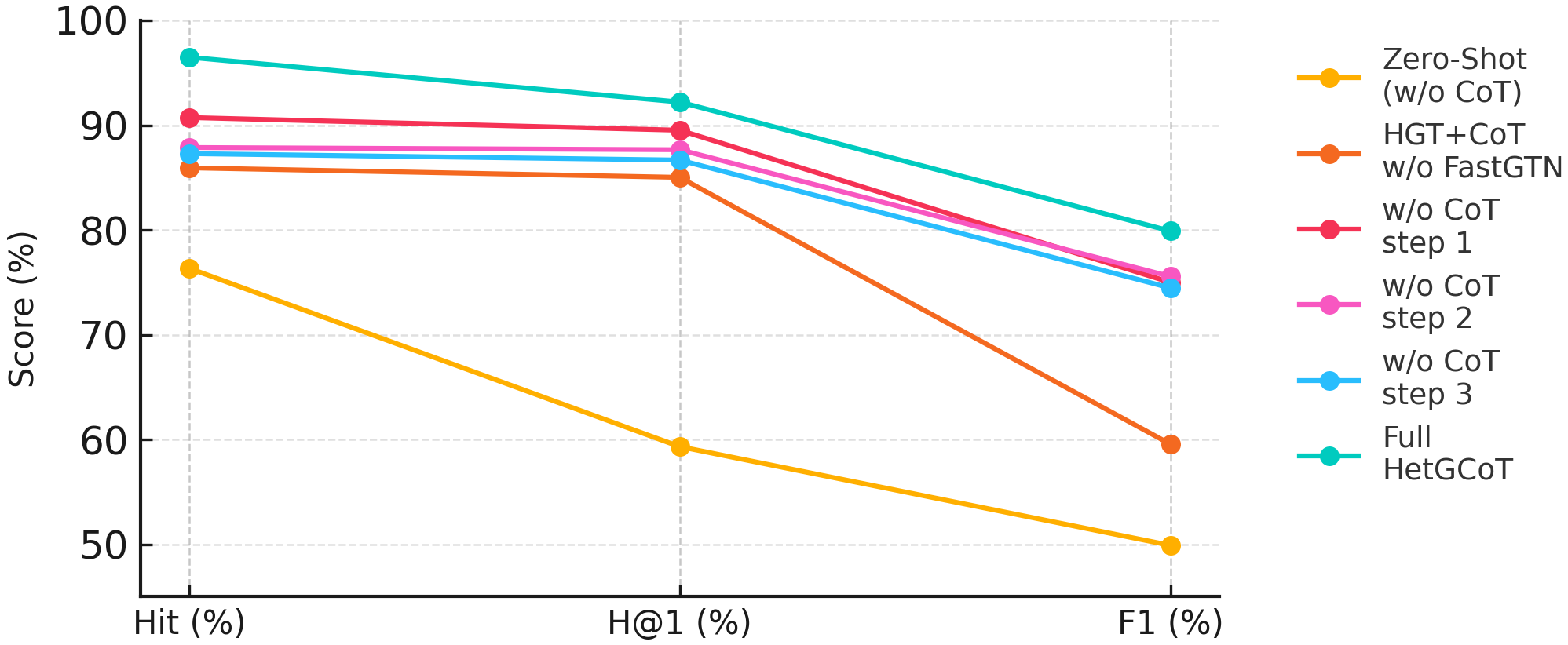}
  \caption{Ablation study of HetGCoT framework components.}
  \label{fig:ablation}
\end{figure}

To assess the contribution of each component in our framework, we conducted a series of ablation experiments on the Openalex Dataset, as shown in Figure~\ref{fig:ablation}.

The ablation study reveals the importance of each component in our framework. Comparing the zero-shot baseline with the HGT+CoT variant shows that incorporating chain-of-thought reasoning yields substantial performance gains. Further analysis of individual reasoning steps indicates that each contributes to the final performance, with step 3 (collaboration relationships) showing the most impact when removed. The complete HetGCoT framework outperforms all partial configurations, indicating that the four-step reasoning process works synergistically. 

These results validate our design rationale: effectively integrating heterogeneous graph information with each step of the chain-of-thought reasoning process can significantly enhance academic journal recommendation performance.

\section{Conclusion}

In this work, we proposed HetGCoT, a framework that integrates heterogeneous graph neural networks with large language models for academic question answering. Our framework introduces three main contributions: (1) a unified framework that transforms heterogeneous graph structural information into natural language reasoning chains, (2) an adaptive metapath selection mechanism that identifies relevant subgraphs for academic queries, and (3) a multi-step reasoning strategy that incorporates graph-derived contexts into chain-of-thought prompting. Through comprehensive experiments on OpenAlex and DBLP datasets, we demonstrated that HetGCoT significantly outperforms baseline methods. We also validated the framework's adaptability across different LLM architectures. Future work could extend this approach to more complex academic reasoning tasks, incorporate additional relationship types in scholarly networks, and scale to larger interdisciplinary datasets. The combination of structural graph information with language model reasoning presents promising directions for academic information processing systems.

\section*{Limitations}

While our work demonstrates the effectiveness of integrating heterogeneous graph neural networks with LLMs, several limitations should be acknowledged. Firstly, the LLM outputs exhibit some instability across different runs, particularly for complex queries requiring multi-step reasoning. Although we fine-tuned the LLMs to improve stability, future work could explore more robust methods for ensuring consistent reasoning paths. Additionally, the computational requirements for processing large heterogeneous academic graphs remain considerable, potentially limiting real-time applications without further optimization.

\section*{Acknowledgments}
This work was supported by the Commonwealth Scientific and Industrial Research Organization (CSIRO), Australia, in conjunction with the National Science Foundation (NSF) of the United States, under grant CSIRO-NSF \#2303037.

\bibliography{custom}

\appendix

\section{Appendix}
\label{sec:appendix}

\subsection{Experiment Setup Detail}

\paragraph{Dataset Detail}
We conduct experiments using two publicly available academic datasets: OpenAlex and DBLP.

\textbf{OpenAlex License.} OpenAlex is released under the \textit{Creative Commons Attribution 4.0 International License (CC BY 4.0)}. This license allows for reuse, redistribution, and modification, provided that proper attribution is given. More information is available at \url{https://creativecommons.org/licenses/by/4.0/}.

\textbf{DBLP License.} The DBLP computer science bibliography is provided under the \textit{Open Data Commons Attribution License (ODC-BY 1.0)}. This license permits use, sharing, and adaptation of the dataset with attribution. Details are available at \url{https://opendatacommons.org/licenses/by/1.0/}.

Our dataset consists of three primary node types—papers, venues, and authors—each with distinct attribute sets as detailed in Table~\ref{tab:node_attributes}. The graph structure captures the relationships between these entities through directed edges. We split the data into training and test sets in a 9:1 ratio.

\begin{table}[h]
  \centering
  \begin{tabular}{ll}
    \hline
    \textbf{Node Type} & \textbf{Attribute Set} \\
    \hline
    paper  & type, title, year, cited\_count, \\
           & fwci, keywords, abstract \\
    venue  & type, name \\
    author & type, name, organization \\
    \hline
  \end{tabular}
  \caption{Node types and their attribute sets.}
  \label{tab:node_attributes}
\end{table}

The complete node and edge templates follow this structure:

\begin{quote}
\small
\textbf{Node Templates:}
\begin{itemize}
\item \textit{Paper:} ID: \texttt{<paper\_id>}, Attributes: \{type, title, year, cited\_count, fwci, keywords[], abstract\}
\item \textit{Venue:} ID: \texttt{<venue\_id>}, Attributes: \{type, name\}
\item \textit{Author:} ID: \texttt{<author\_id>}, Attributes: \{type, name, organization\}
\end{itemize}

\textbf{Edge Templates:}
\begin{itemize}
\item \textit{Paper-venue:} (src: \texttt{<paper\_id>}, dst: \texttt{<venue\_id>}), Attributes: \{type: 'paper-venue'\}
\item \textit{Paper-author:} (src: \texttt{<paper\_id>}, dst: \texttt{<author\_id>}), Attributes: \{type: 'paper-author'\}
\end{itemize}
\end{quote}

\paragraph{Implementation Detail}
For Sentence-BERT, We obtain a single 768-dimensional embedding per node by concatenating its title, abstract, and keywords into a Sentence-BERT model. This 768-dimensional vector is then augmented with two numeric attributes (total citation count and FWCI) to form a 770-dimensional feature vector for each paper node.

For HGT, We feed the 770-dimensional feature into a two-layer Heterogeneous Graph Transformer (HGT) with type-specific Q-K-V projections, 8 attention heads, and a hidden size of $d=387$. The model is trained for 100 epochs using Adam (learning rate $1\times10^{-3}$) on the paper–venue link‐prediction task. Final per‐node embeddings ($h_i\in\mathbb R^{387}$) are saved for downstream use.

For FastGTN, a lightweight FastGTN autoencoder (7 layers, 8 channels, hidden size 64) is trained to reconstruct the frozen HGT embeddings (minimizing MSE$(Z_{\text{paper}},h_{\text{paper}})$) over 50 epochs with Adam (learning rate $3\times10^{-3}$) on GPU. After training, we extract the learned relation‐importance weights and score each candidate metapath by summing its edge weights.

\subsection{Statistical Robustness Analysis}
To verify the statistical robustness of our approach, we conducted three independent runs with different random seeds on both datasets. Table~\ref{tab:statistical_robustness} presents the detailed results across all metrics.

\begin{table}[h!]
\caption{Statistical robustness analysis across three independent runs with different random seeds.}
\label{tab:statistical_robustness}
\centering
\scriptsize
\setlength{\tabcolsep}{2pt}
\begin{tabular}{lccccc}
\toprule
\textbf{Dataset} & \textbf{Run} & \textbf{Hit} & \textbf{H@1} & \textbf{F1} & \textbf{NDCG} \\
\midrule
\multirow{4}{*}{OpenAlex} 
& Run 1 & 95.73 & 91.45 & 82.91 & 91.29 \\
& Run 2 & 96.48 & 92.21 & 79.90 & 91.21 \\
& Run 3 & 96.50 & 92.00 & 78.00 & 92.00 \\
\cmidrule{2-6}
& \textbf{Mean$\pm$Std} & \textbf{96.24$\pm$0.36} & \textbf{91.89$\pm$0.32} & \textbf{80.27$\pm$2.02} & \textbf{91.50$\pm$0.36} \\
\midrule
\multirow{4}{*}{DBLP}
& Run 1 & 85.31 & 83.70 & 64.55 & 83.49 \\
& Run 2 & 84.70 & 83.60 & 64.66 & 83.39 \\
& Run 3 & 85.27 & 83.75 & 64.38 & 83.46 \\
\cmidrule{2-6}
& \textbf{Mean$\pm$Std} & \textbf{85.09$\pm$0.28} & \textbf{83.68$\pm$0.06} & \textbf{64.53$\pm$0.12} & \textbf{83.45$\pm$0.04} \\
\bottomrule
\end{tabular}
\end{table}

The results demonstrate consistent performance across multiple runs. This consistency confirms the stability of our approach across different experimental conditions.

\subsection{Evaluation with Reasoning-Type LLMs}
To assess the compatibility of our framework with advanced reasoning-type language models, we conducted additional experiments using OpenAI o3 and DeepSeek-R1 without fine-tuning. Table~\ref{tab:reasoning_llms} presents the comparative results.

\begin{table}[h!]
\caption{Performance comparison with reasoning-type LLMs.}
\label{tab:reasoning_llms}
\centering
\scriptsize
\setlength{\tabcolsep}{2pt}
\begin{tabular}{lccccc}
\toprule
\textbf{Dataset} & \textbf{Model} & \textbf{Hit} & \textbf{H@1} & \textbf{F1} & \textbf{NDCG} \\
\midrule
\multirow{5}{*}{OpenAlex}
& o3 & 74.82 & 59.46 & 39.56 & 65.38 \\
& o3+HetGCoT & 90.90 & 86.18 & 68.75 & 85.46 \\
\cmidrule{2-6}
& DeepSeek-R1 & 73.58 & 60.38 & 41.51 & 62.26 \\
& DeepSeek-R1+HetGCoT & 92.16 & 88.24 & 74.51 & 78.43 \\
\cmidrule{2-6}
& \textbf{HetGCoT (ours)} & \textbf{96.48} & \textbf{92.21} & \textbf{79.90} & \textbf{91.29} \\
\midrule
\multirow{5}{*}{DBLP}
& o3 & 62.58 & 51.67 & 41.60 & 60.08 \\
& o3+HetGCoT & 78.14 & 68.62 & 53.54 & 67.83 \\
\cmidrule{2-6}
& DeepSeek-R1 & 63.16 & 57.89 & 43.86 & 59.65 \\
& DeepSeek-R1+HetGCoT & 79.60 & 71.43 & 61.22 & 69.39 \\
\cmidrule{2-6}
& \textbf{HetGCoT (ours)} & \textbf{85.31} & \textbf{83.70} & \textbf{64.55} & \textbf{83.49} \\
\bottomrule
\end{tabular}
\end{table}

The results demonstrate that our HetGCoT framework maintains its effectiveness when applied to reasoning-type language models, with consistent performance improvements observed across different model architectures.

\subsection{Ablation Study}
The details of our ablation study are shown in Table~\ref{tab:ablation}.

\begin{table}[H]
  \centering
  \setlength{\tabcolsep}{4pt}
  \resizebox{\columnwidth}{!}{%
    \begin{tabular}{lccc}
      \toprule
      \textbf{Variant} & \textbf{Hit (\%)} & \textbf{H@1 (\%)} & \textbf{F1 (\%)} \\
      \midrule
      Zero-Shot (w/o CoT)          & 76.34 & 59.32 & 49.90 \\
      HGT+CoT w/o FastGTN          & 85.93 & 85.02 & 59.56 \\
      w/o CoT reasoning step 1     & 90.73 & 89.52 & 75.00 \\
      w/o CoT reasoning step 2     & 87.87 & 87.66 & 75.56 \\
      w/o CoT reasoning step 3     & 87.29 & 86.67 & 74.47 \\
      \textbf{Full HetGCoT}        & \textbf{96.48} & \textbf{92.21} & \textbf{79.90} \\
      \bottomrule
    \end{tabular}%
  }
  \caption{Ablation study of HetGCoT framework components}
  \label{tab:ablation}
\end{table}

\subsection{Algorithm}
The pseudocode of our method is shown in Table~\ref{tab:code}, and all experiments were conducted on two A100 GPUs.

\begin{table}[H]
  \centering
  \small                           
  \setlength{\tabcolsep}{4pt}      
  \resizebox{\columnwidth}{!}{
    \begin{tabular}{l}
      \toprule
      \textbf{Algorithm 1} HetGCoT algorithm. $G$ is a heterogeneous graph, $q$ is the query, $K$ is the number of \\
      selected metapaths per template and $\gamma$ is the length normalization factor \\
      \midrule
      1: \textbf{procedure} HetGCoT($G, q, K, \gamma$) \\
      2: \quad $H \leftarrow \text{HGT-ENCODE}(G)$ \hfill $\triangleright$ Heterogeneous graph construction \\
      3: \quad $W \leftarrow \text{FASTGTN-TRAIN}(G, H)$ \\
      4: \quad $S_q \leftarrow \text{COSINE-SIMILARITY}(H_q, H)$ \\
      5: \quad $\text{MP}_{\text{candidates}} \leftarrow \text{GENERATE-METAPATHS}(G, S_q)$ \hfill $\triangleright$ HGT Metapath Pool Generation \\
      6: \quad \textbf{for} $\text{mp} \in \text{MP}_{\text{candidates}}$ \textbf{do} \\
      7: \quad \quad $\text{scores}[\text{mp}] \leftarrow \sum_{(u,v) \in \text{mp}} W[\psi(u,v)] / |\text{mp}|^\gamma$ \hfill $\triangleright$ Adaptive Metapaths Selection \\
      8: \quad \textbf{end for} \\
      9: \quad $\text{MP}_{\text{selected}} \leftarrow \text{STRATIFIED-TOP-K}(\text{scores}, K)$ \\
      10: \quad \textbf{for} $\text{mp} \in \text{MP}_{\text{selected}}$ \textbf{do} \\
      11: \quad \quad $\text{contexts} \leftarrow \text{contexts} \cup \text{NATURALIZE}(\text{mp}, \text{scores}[\text{mp}])$ \\
      12: \quad \textbf{end for} \\
      13: \quad $\text{prompt} \leftarrow \text{CONSTRUCT-PROMPT}(q, \text{contexts})$ \\
      14: \quad $\text{answer} \leftarrow \text{LLM-COT-REASONING}(\text{prompt})$ \hfill $\triangleright$ 4-step reasoning \\
      15: \quad \textbf{return} $\text{answer}$ \\
      16: \textbf{end procedure} \\
      \bottomrule
    \end{tabular}%
  }
  \caption{HetGCoT algorithm.}
  \label{tab:code}
\end{table}

\subsection{Prompt Templates}

\begin{table*}[t]
  \centering
  \small
  \begin{tabular}{p{0.95\textwidth}}
    \toprule
    \textbf{System Prompt} \\ 
    \midrule
    You are a professional academic journal recommendation expert. Your task is to recommend the three most suitable journals for publishing the provided paper information, following the specified reasoning steps, and to explain the reasons for each recommendation in detail. Please note: \\
    $\bullet$ Each paper can have only one correct publishing journal, which should be placed at the top of the recommendation list. \\
    $\bullet$ Please strictly follow the reasoning steps below and use the provided specific related information to answer within the given journal list. \\
    \midrule
    \textbf{User Prompt} \\
    \midrule
    Please recommend the three most suitable journals for publishing this paper based on the following information, strictly following the specified reasoning steps, and explain the reasons. \\[4pt]
    \textbf{Step 1: Learn the graph structure information related to each journal based on the following predefined metapaths} \\[2pt]
    \textit{Question:} Based on the following predefined metapaths, learn the graph structure information related to each journal. \\[2pt]
    \textit{Provided Information:} \\[2pt]
    [Metapaths] \\
    APVPA Metapaths with confidence score: \\
    APVPA Metapath 1 \\
    APVPA Metapath 2 \\
    APVPA Metapath 3 \\
    APVPA Metapath 4 \\
    APVPA Metapath 5 \\[2pt]
    VPAPV Metapaths with confidence score: \\
    VPAPV Metapath 1 \\
    VPAPV Metapath 2 \\
    VPAPV Metapath 3 \\
    VPAPV Metapath 4 \\
    VPAPV Metapath 5 \\[2pt]
    --- \\[2pt]
    \textbf{Step 2: Identify the core themes and keywords of the paper, and define the paper's research field} \\[2pt]
    \textit{Question:} Based on the following paper description, identify the core themes and keywords of the paper, its impact level, and determine its research field. \\[2pt]
    \textit{Provided Information:} \\[2pt]
    [Paper Description] \\
    Paper [ ], titled [ ], has [ ] citations, FWCI (Field-Weighted Citation Impact) of [ ], authored by [ ], published in [ ], topics: [ ], abstract: [ ] \\
    --- \\[2pt]
    \textbf{Step 3: Analyze the collaboration information of the authors} \\[2pt]
    \textit{Question:} Based on the following collaboration relationships, analyze the collaboration status between the authors, their common research directions, and joint publications. \\[2pt]
    \textit{Provided Information:} \\[2pt]
    [Collaboration Relationships] \\
    Author [ ] is affiliated with [ ], mainly publishes papers in journals such as … \\
    Author [ ] is affiliated with [ ], mainly publishes papers in journals such as … \\[2pt]
    --- \\[2pt]
    \textbf{Step 4: Based on the above information, recommend the three most suitable journals for publishing this paper from the journal list below, sorted by probability from high to low, and explain the reasons} \\[2pt]
    \textit{Question:} Based on the above analysis of the paper's content, authors' backgrounds, collaboration relationships, and the learned graph structure information, recommend the three most suitable journals for publishing this paper from the journal list below, sorted by probability from high to low, and provide detailed reasons for each recommendation. \\[2pt]
    \textit{Provided Information:} \\[2pt]
    [Journal List] \\
    - 1. … \\
    - 2. … \\
    - 3. … \\
    - 4. … \\
    - 5. … \\
    \midrule
    \textbf{Assistant Response Format} \\
    \midrule
    Recommended Journals: \\
    1. … \\
    2. … \\
    3. … \\
    Detailed explanation according to the reasoning procedure \\
    \bottomrule
  \end{tabular}
  \caption{Prompt template for journal recommendation.}
  \label{tab:journal_template}
\end{table*}

\begin{table*}[t]
  \centering
  \small
  \begin{tabular}{p{0.95\textwidth}}
    \toprule
    \textbf{System Prompt} \\
    \midrule
    You are an academic graph-reasoning assistant. Your task is to analyze paper authorship patterns to predict which papers are most likely written by a specific author. Please strictly follow the provided reasoning steps and use the provided graph structure information and paper context to make your predictions. \\
    $\bullet$ Each query focuses on one target author and requires selecting the three most likely papers from five candidates. \\
    $\bullet$ Please strictly follow the reasoning steps below and use the provided metapath information and paper descriptions. \\
    \midrule
    \textbf{User Prompt} \\
    \midrule
    Please identify the three most likely papers written by the target author based on the following information, strictly following the specified reasoning steps. \\[4pt]
    \textbf{Step 1: Graph Structure via Metapaths} \\[2pt]
    \textit{Question:} Based on the following metapaths, learn the academic heterogeneous graph structure and relationships. \\[2pt]
    \textit{Provided Information:} \\[2pt]
    [Metapaths] \\
    APVPA Metapaths with confidence score: \\
    APVPA Metapath 1 \\
    APVPA Metapath 2 \\
    APVPA Metapath 3 \\
    APVPA Metapath 4 \\
    APVPA Metapath 5 \\[2pt]
    VPAPV Metapaths with confidence score: \\
    VPAPV Metapath 1 \\
    VPAPV Metapath 2 \\
    VPAPV Metapath 3 \\
    VPAPV Metapath 4 \\
    VPAPV Metapath 5 \\[2pt]
    --- \\[2pt]
    \textbf{Step 2: Paper Context} \\[2pt]
    \textit{Question:} Based on the following paper description, understand the research content, themes, and academic context. \\[2pt]
    \textit{Provided Information:} \\[2pt]
    [Paper Description] \\
    Paper [ ], titled [ ], has [ ] citations, FWCI (Field-Weighted Citation Impact) of [ ], published in [ ], topics: [ ], abstract: [ ] \\
    --- \\[2pt]
    \textbf{Step 3: Make Prediction} \\[2pt]
    \textit{Question:} Based on the above analysis of the graph structure and paper context, choose the three most likely papers written by author \{author\} from the paper list below, sorted by probability from high to low. \\[2pt]
    \textit{Provided Information:} \\[2pt]
    [Paper List] \\
    - \{paper\_1\} \\
    - \{paper\_2\} \\
    - \{paper\_3\} \\
    - \{paper\_4\} \\
    - \{paper\_5\} \\[2pt]
    \midrule
    Based on the analysis, the three most likely papers are: \\
    1. \{paper\_1\} \\
    2. \{paper\_2\} \\
    3. \{paper\_3\} \\
    Detailed explanation according to the reasoning procedure \\
    \bottomrule
  \end{tabular}
  \caption{Prompt template for authorship identification (paper-
author relationships).}
  \label{tab:author_paper_template}
\end{table*}

\begin{table*}[t]
  \centering
  \small
  \begin{tabular}{p{0.95\textwidth}}
    \toprule
    \textbf{System Prompt} \\
    \midrule
    You are an academic graph‐reasoning assistant. Your task is to analyze academic collaboration networks to predict which researchers are most likely to be collaborators of a specific author. Please strictly follow the provided reasoning steps and use the provided graph structure information and paper context to make your predictions. \\
    $\bullet$ Each query focuses on one target author and requires selecting the three most likely collaborators from five candidates. \\
    $\bullet$ Please strictly follow the reasoning steps below and use the provided metapath information and paper descriptions. \\
    \midrule
    \textbf{User Prompt} \\
    \midrule
    Please identify the three most likely collaborators of the target author based on the following information, strictly following the specified reasoning steps. \\[4pt]
    \textbf{Step 1: Graph Structure via Metapaths} \\[2pt]
    \textit{Question:} Based on the following metapaths, learn the academic heterogeneous graph structure and author relationships. \\[2pt]
    \textit{Provided Information:} \\[2pt]
    [Metapaths] \\
    APVPA Metapaths with confidence score: \\
    APVPA Metapath 1 \\
    APVPA Metapath 2 \\
    APVPA Metapath 3 \\
    APVPA Metapath 4 \\
    APVPA Metapath 5 \\[2pt]
    VPAPV Metapaths with confidence score: \\
    VPAPV Metapath 1 \\
    VPAPV Metapath 2 \\
    VPAPV Metapath 3 \\
    VPAPV Metapath 4 \\
    VPAPV Metapath 5 \\[2pt]
    --- \\[2pt]
    \textbf{Step 2: Paper Context} \\[2pt]
    \textit{Question:} Based on the following paper description, understand the research domain and collaboration context. \\[2pt]
    \textit{Provided Information:} \\[2pt]
    [Paper Description] \\
    Paper [ ], titled [ ], has [ ] citations, FWCI (Field-Weighted Citation Impact) of [ ], published in [ ], topics: [ ], abstract: [ ] \\
    --- \\[2pt]
    \textbf{Step 3: Make Prediction} \\[2pt]
    \textit{Question:} Based on the above analysis of the graph structure and research context, choose the three most likely collaborators of author \{author\_id\} from the researcher list below, sorted by probability from high to low. \\[2pt]
    \textit{Provided Information:} \\[2pt]
    [Author List] \\
    - \{author\_1\} \\
    - \{author\_2\} \\
    - \{author\_3\} \\
    - \{author\_4\} \\
    - \{author\_5\} \\[2pt]
    \midrule
    Based on the analysis, the three most likely collaborators are: \\
    1. \{author\_1\} \\
    2. \{author\_2\} \\
    3. \{author\_3\} \\
    Detailed explanation according to the reasoning procedure \\
    \bottomrule
  \end{tabular}
  \caption{Prompt template for collaboration discovery (author-author relationships).}
  \label{tab:author_collab_template}
\end{table*}

\end{document}